\documentclass[twocolumn,preprintnumbers,amsmath,amssymb]{revtex4}

\usepackage{graphicx}
\usepackage{dcolumn}
\usepackage{bm}
\usepackage{epsfig}
\usepackage{amssymb}
\begin{document}

\preprint{}

\title{Comments on ``First Dark Matter Results from the XENON100 
Experiment''}

\author{J.I. 
Collar$^{a}$
}
\author{D.N. McKinsey$^{b}$
}
\address{ 
$^{a}$Enrico 
Fermi Institute, KICP and  Department of Physics, University of Chicago, Chicago, IL 60637\\
}
\address{ 
$^{b}$Department of Physics, Yale University, New Haven, CT 06520\\
}
\maketitle

The XENON100 collaboration has recently released new dark matter 
limits \cite{xenon100}, 
placing particular emphasis on their impact on searches known to be sensitive 
to light-mass (below $\sim$10 GeV/c$^{2}$) Weakly Interacting 
Massive Particles (WIMPs), 
such as DAMA \cite{DAMA} and CoGeNT \cite{cogent}. We describe here several sources of uncertainty 
and bias in their analysis that make their new claimed sensitivity 
presently untenable. In particular, we point out additional work in this field and 
simple kinematic arguments that indicate that liquid xenon (LXe) may be a 
relatively insensitive detection medium for the recoil energies (few 
keV$_{r}$) expected from such low mass WIMPs. To place the 
discussion that follows in some perspective, 
using the most recently suggested mean value of the galactic escape 
velocity \cite{escape}, 
an example 7 GeV/c$^{2}$ WIMP can impart an absolute maximum of 4 keV$_{r}$ to a xenon nucleus, 
with the majority ($\sim$90\%) of the events depositing energies 
below 1.5 keV$_{r}$. 

It is suggested in \cite{xenon100} that the value of $\mathcal{L}_{\text{eff}}$ (the ratio between 
electron equivalent energy and nuclear recoil energy) adopted to 
obtain WIMP limits is constant ($\mathcal{L}_{\text{eff}}\sim$0.12 
below $\sim$10 keV$_{r}$)  
and a representative compromise encompassing all existing low-energy 
measurements for LXe $^{1}$\footnotetext[1]{In a fascinating attempt at 
trompe l'oeil, an arbitrary  monotonically
decreasing $\mathcal{L}_{\text{eff}}$ is plotted in Fig.\ 1 in
\cite{xenon100}, but is nowhere else used. Given the lack of agreement
between the closing paragraph, the Fig.\ 1 caption and the phrase
immediately above Fig.\ 2, we are led to believe that the lower 90\%
contour is missing from Fig.\ 1 below 5 keVr. This can mislead the reader
to identify this contour with the extrapolation to Manzur {\it et 
al.} shown in that figure. In version 2, without truly clarifying 
this standing issue, the authors
now explicitly mention the effect of a
decreasing $\mathcal{L}_{\text{eff}}$ below 5 keVr, 
resulting in much less conflict with DAMA and CoGeNT. We illustrate 
exactly by how much in Fig.\ 5 here.}. 
Nothing is further from reality. 

Attempts to measure 
$\mathcal{L}_{\text{eff}}$ can be classified into 
two methods \cite{dan1}, fixed-energy neutron experiments with scattered neutron tagging like 
those exclusively considered by XENON100, and direct comparisons between broad-spectrum 
neutron source calibration data and a variety of Monte Carlo simulations, like 
those adopted by the ZEPLIN collaboration \cite{zeplin}. Results from 
the latter are included as the red band in Fig.\ 1 here,
displaying the dramatic drop in $\mathcal{L}_{\text{eff}}$ 
observed at recoil energies that would make a LXe search 
for light WIMPs a futile exercise. Interestingly, a drop in response to low 
energy recoils 
seems to be a common feature to 
all other attempts to use the second method (Fig.\ 2) \cite{theiraps}, 
including the most recent by the XENON10 collaboration
\cite{peter}. This last, not shown in these figures, is a reanalysis 
of \cite{prevnim}. None of this is mentioned 
in \cite{xenon100}, where the authors repeatedly refer to their 
selective list of measurements as ``all data''. 

\begin{figure}
\includegraphics[width=8.5cm]{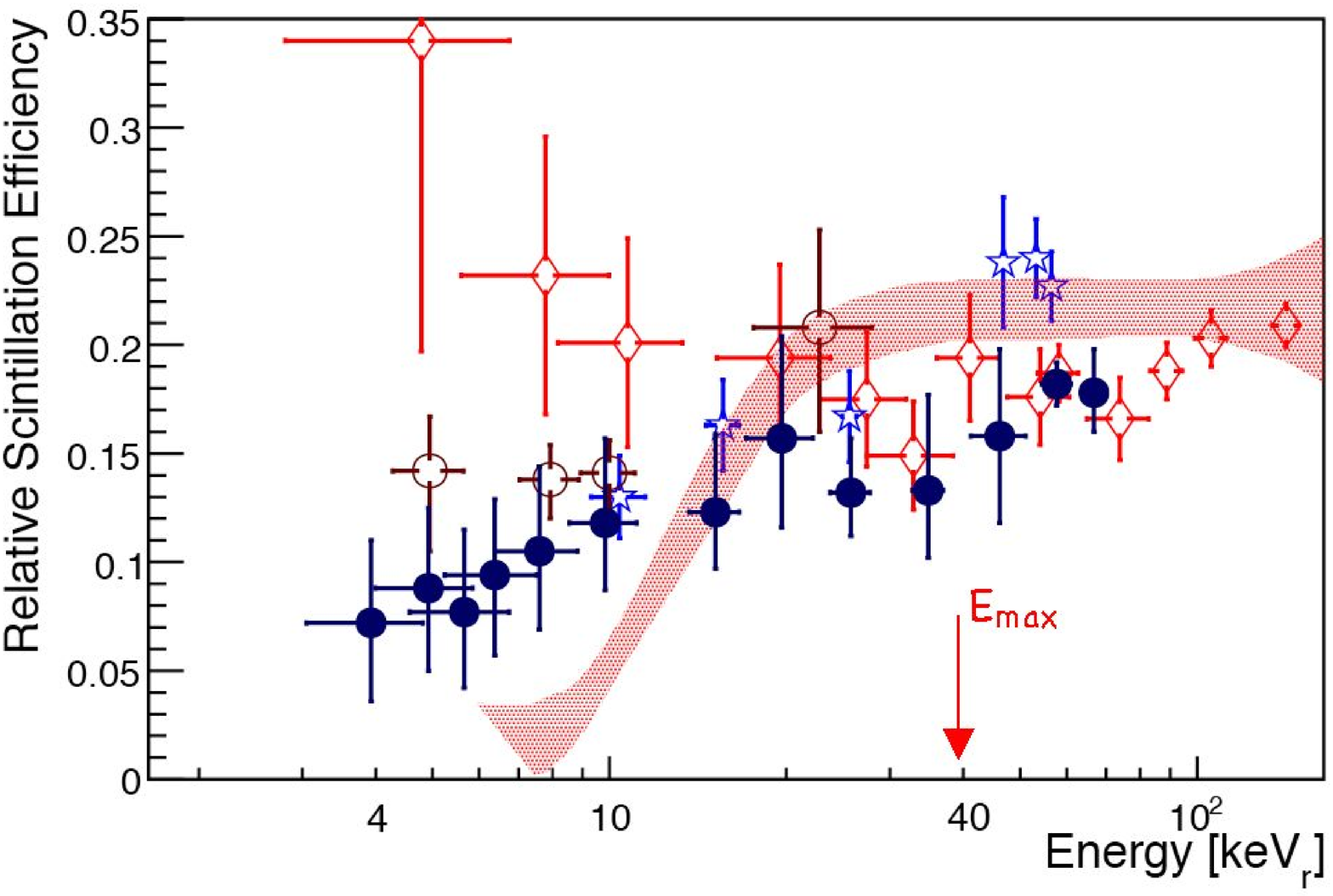}
\caption{\label{fig:epsart}Measurements of $\mathcal{L}_{\text{eff}}$ 
in LXe. The red vertical arrow indicates the calculated value for the 
kinematic cutoff in recoil energy (see text). The most recent 
analysis by the XENON10 collaboration \protect\cite{peter}, not considered in 
\protect\cite{xenon100}, follows the same trend as in Manzur {\it et al.} \protect\cite{dan1} (dark blue 
points here). Light-mass WIMPs like those claimed to be excluded in 
\protect\cite{xenon100} concentrate their signal beyond the left 
margin of this figure. A constant $\mathcal{L}_{\text{eff}}\sim$0.12 below $\sim$10 keV$_{r}$
is used in \protect\cite{xenon100} to obtain dark matter limits.
}
\end{figure}

\begin{figure}
\includegraphics[width=9cm]{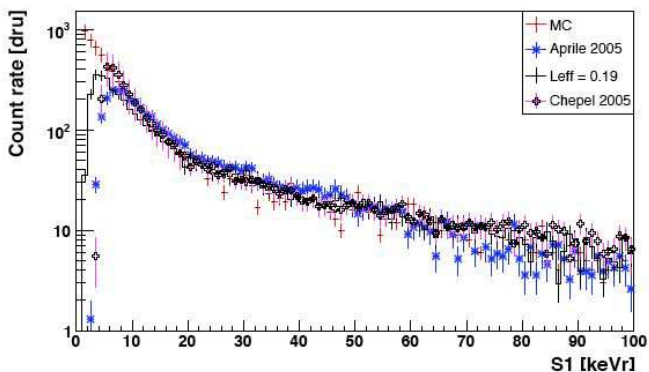}
\caption{\label{fig:epsart}Examples of comparisons between simulation 
(MC) and response of LXe chambers to neutron sources 
\protect\cite{theiraps} (see text). The dramatic disagreement at few 
keV$_{r}$ can be alleviated to some extent via trigger efficiency 
corrections, but not completely \protect\cite{peter}.
}
\end{figure}

While virtues and defects can be listed for both methods,
a common feature of most of these measurements is the value of $\mathcal{L}_{\text{eff}}$ 
(few tens of keV$_{r}$) at which the drastic drop in recoil sensitivity appears. 
This onset and ensuing trend is also visible in the data from \cite{dan1}, 
the most recent fixed-energy experiment, featuring the best control 
of systematics so far 
for that particular family of $\mathcal{L}_{\text{eff}}$ 
measurements. Historically speaking, the evolution of fixed-energy 
experiments has proceeded monotonically towards pointing to modest recoil response at the lowest 
energies, i.e., towards reconciliation with broad-spectrum calibrations. 
This is a fact hidden from view in \cite{xenon100} and that clashes 
with their choice of $\mathcal{L}_{\text{eff}}$ (Fig.\ 1). Most researchers 
in the field will argue that 
this behavior (null $\mathcal{L}_{\text{eff}}$ at zero or small 
recoil energy) is to be naturally expected. 
Next we provide at least one physical mechanism supporting this.

The marked drop in 
$\mathcal{L}_{\text{eff}}$ at low energies in the experiments that the XENON100 
collaboration has ignored may be understood from simple two-body kinematics 
affecting the energy transfer from a xenon recoil to an atomic 
electron. As already discussed
within the context of the MACRO experiment \cite{ahlen}, a kinematic cutoff to the 
production of scintillation is expected whenever the minimum 
excitation energy E$_{g}$  
of the system exceeds the maximum possible energy transfer to an electron 
by a slow-moving recoil (E$_{max}$). Unfortunately, such a basic consideration is often 
not included in attempts to develop an understanding of scintillator 
response to very slow 
ions \cite{review}, but this is not always the case \cite{spooner}. 
In practice, and for the reasons described in \cite{ahlen}, 
a smooth adiabatic drop is observed rather than a step-like cutoff. While it is widely 
acknowledged that much is left to be understood about the exact mechanisms leading to 
scintillation from low-energy ions in LXe (per se an obvious reason to act 
very conservatively), direct atomic excitation is known to 
compete with the recombination of ionized electrons 
in producing scintillation. Even then, the initial ratio of excitons to electron-ion pairs
is estimated by a faction of workers in this field to be small at 
0.06-0.2 \cite{dan1,doke}. 
It would then be reasonable to identify E$_{g}$ with the band gap in LXe (9.3 eV), 
and to expect this cutoff and the smooth decrease in $\mathcal{L}_{\text{eff}}$ below it to appear at
a calculated E$_{max}\sim$ 39 keV$_{r}$ for LXe. 
We remark that this value is in good agreement with the behavior 
noticed in the family of $\mathcal{L}_{\text{eff}}$ measurements 
disregarded by XENON100 (Figs. 1,2) and 
with the evident trend most recently observed in 
\cite{dan1,peter,aprile}.

Fig.\ 3 displays approximate values for E$_{max}$ for other scintillators used in 
dark matter research, calculated following \cite{ahlen,ziegler,lumin}. While the 
measurements of quenching factor or relative scintillation efficiency 
(similar in meaning to $\mathcal{L}_{\text{eff}}$ for purposes of 
this discussion) displayed in Fig.\ 3 are just a few representative examples, 
they span the range of energies explored in recoil calibrations up 
until now \cite{review}. The comparatively large 
value of E$_{max}$ for LXe is a result of its large recoil mass and 
relatively high E$_{g}$. 
It is interesting to observe that the recoil energy range well below E$_{max}$ has only 
been experimentally explored for LXe and plastic scintillator 
(NE-110) \cite{ahlen}, and that in 
both instances the expected drop in sensitivity is readily observed at the predicted 
onset. This brings up a parallel observation that the low-energy trend in quenching factor for 
recoils in NaI[Tl]  scintillators is also an experimental unknown: if 
the quenching factor rises before reaching E$_{max}$, 
a expected behavior \cite{review} observed for the materials in Fig.\ 3 and for liquid 
scintillator \cite{spooner}, 
this could have important implications for light WIMP interpretations of the DAMA annual modulation, 
and in particular its agreement or not with CoGeNT. 

\begin{figure}
\includegraphics[width=9cm]{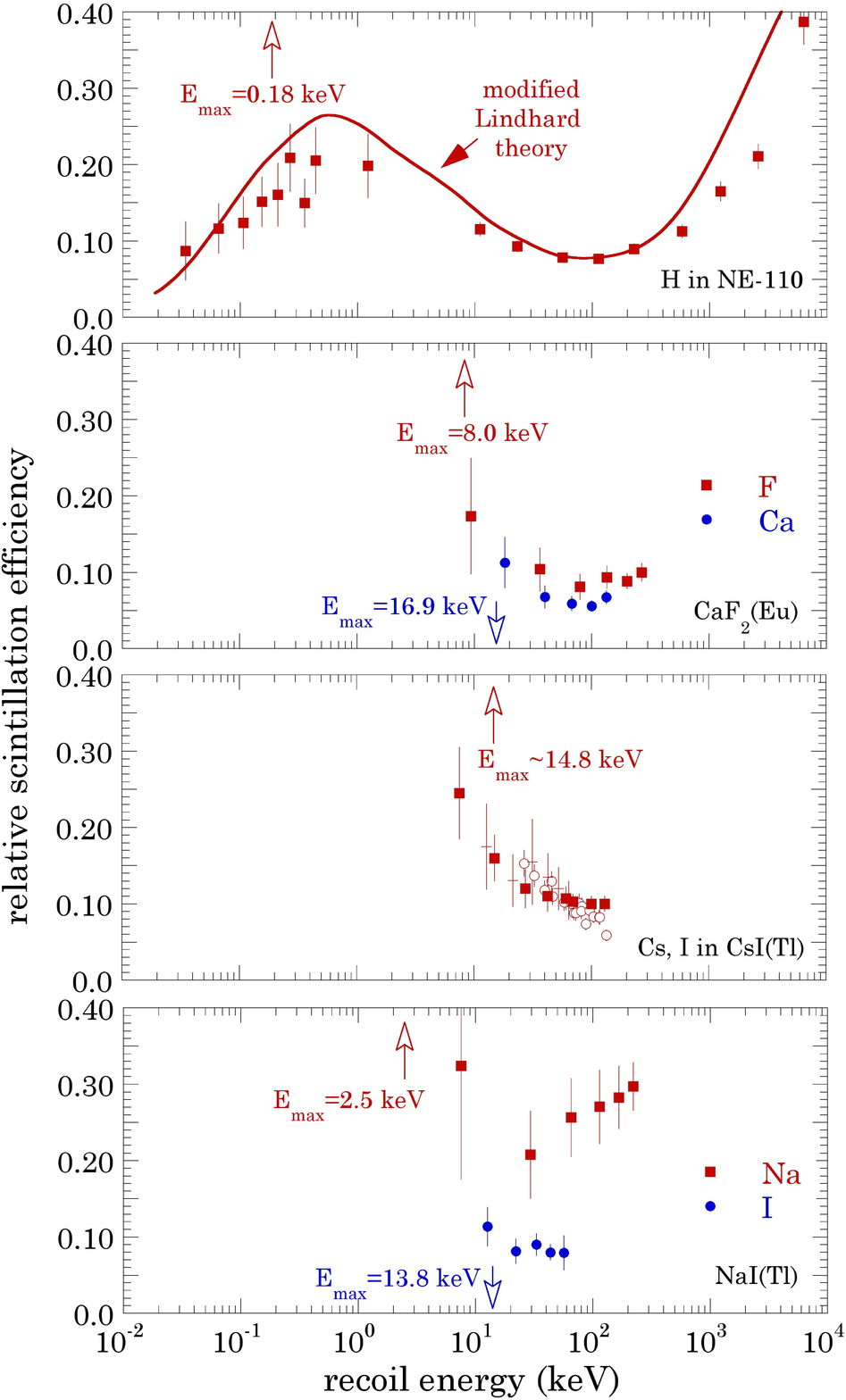}
\caption{\label{fig:epsart}Examples of measurements of quenching 
factor for different scintillators, indicating in each case  the calculated value for the 
kinematic cutoff in recoil energy (see text). An additional value of 
E$_{max}$=2.2 keV$_{r}$ is found for carbon recoils in organic 
scintillator, in good agreement with \protect\cite{spooner}.
}
\end{figure}

There are several other weaknesses in the reasoning leading to the 
claims in \cite{xenon100}. For instance, the new 
claimed XENON100 limits depend 
critically (by several orders of magnitude below $\sim$10 GeV/c$^{2}$) on the assumption of a Poisson tail in the 
modest number of photoelectrons that would be generated by a light-mass WIMP above detection threshold, 
even for their forced choice of $\mathcal{L}_{\text{eff}}$. 
We question the wisdom of this approach when the mechanisms behind the 
generation of {\it any} significant amount of scintillation are still 
unknown and
may simply be absent at the few keV$_{r}$ level. To put it bluntly, 
this is the equivalent of expecting something out of nothing. An 
example of the level of sensitivity expected from XENON100 in the 
absence of this assumption can be found in \cite{wonder}: WIMPs with 
a mass lower than $\sim$12 GeV/c$^{2}$ are then entirely out of reach for 
XENON100, imposing no significant constraints on DAMA, CoGeNT, or any 
other dark matter detector technology with demonstrated sensitivity to this mass 
region. 

It seems clear that sufficient knowledge on the energy dependence of 
$\mathcal{L}_{\text{eff}}$ in the region 0-3 keV$_{r}$
is presently absent for LXe at the excellent 
level that would be required to establish reliable light-WIMP limits. 
This begs comparison with highly linear detecting media such as 
germanium detectors, for which careful 
dedicated measurements of quenching factor have been made down to 
$\sim$0.25 keV$_{r}$, measurements found to be 
in good agreement with theory \cite{jcapetc,tex}.  

\begin{figure}
\includegraphics[width=8cm]{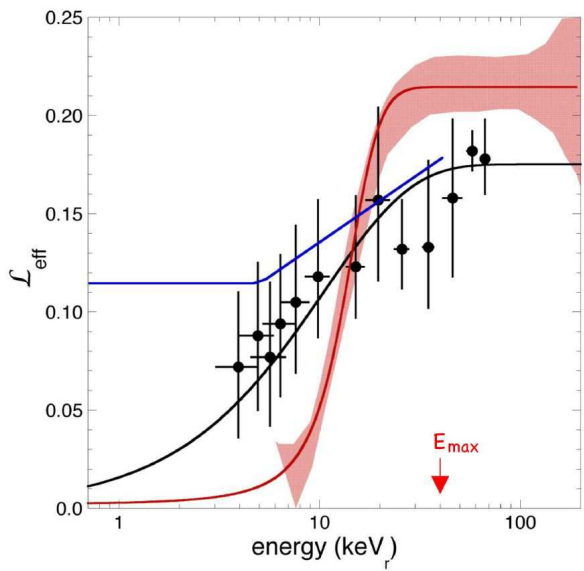}
\caption{\label{fig:epsart}Models for $\mathcal{L}_{\text{eff}}$ used 
to generate the exclusions in Fig.\ 5. The blue line corresponds to 
the extrapolation to low energy used in \protect\cite{xenon100}. 
Black points are recent data 
from Manzur {\it et al.} \protect\cite{dan1}. The black line is an 
adiabatic fit to these (see text). The red line (logistic 
fit) and band correspond to ZEPLIN 
data \protect\cite{zeplin} (see text).
}
\end{figure}
\begin{figure}
\includegraphics[width=8cm]{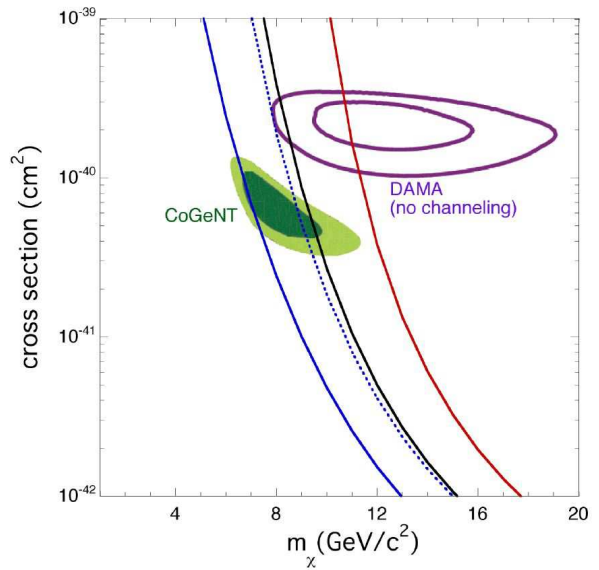}
\caption{\label{fig:epsart} XENON100 exclusions generated from the 
$\mathcal{L}_{\text{eff}}$ models in Fig.\ 4. (same color coding). 
The dotted blue line corresponds to the problematic $\mathcal{L}_{\text{eff}}$
contour in Fig.\ 1 in \protect\cite{xenon100}, ambiguously claimed by 
XENON100 to be both a 90\% lower C.L. to their 
$\mathcal{L}_{\text{eff}}$ best fit 
and an extrapolation to Manzur {\it et al.} \protect\cite{dan1}, a crucial point we contend is 
misleading.
Ion channeling is not included neither for CoGeNT nor 
DAMA regions \protect\cite{andreas,gondolo}. 
The low-energy trend 
expected for the quenching factor in NaI(Tl) \protect\cite{review}, 
not included here, 
can have the effect of displacing the DAMA region away from 
XENON100 constraints (see text). 
}
\end{figure}

In conclusion, we find that the choice made in \cite{xenon100} in relation to the 
low-energy trend for $\mathcal{L}_{\text{eff}}$, a constant value below 
10 keV$_{r}$, is
not only biased (clashing with several ignored experimental measurements 
and the observed historical trend), but also seemingly unphysical, as would be derived from simple kinematic 
considerations. We detect an intent in \cite{xenon100} to avoid considering important standing issues in 
this area of research, as well as serious contradictions around the 
meaning and content of their 
Fig.\ 1. We firmly maintain that the low-mass WIMP limits presented by the XENON100 collaboration are 
the least conservative choice over a present uncertainty spanning several orders of magnitude, 
including the very real possibility that LXe is an effectively inert detection medium 
for WIMPs in this low-mass range. As such, these limits are untenable. 
A more conservative treatment of present-day uncertainties in 
$\mathcal{L}_{\text{eff}}$
(such as for instance, that adopted by the ZEPLIN collaboration 
\cite{zeplin}) would also 
lead to weaker limits at higher WIMP masses, raising an additional 
question on the relevance of the present XENON100 
sensitivity in comparison to that from CDMS and other experiments. 

The onus of unequivocally demonstrating the 
existence of scintillation light from $\sim$1 keV$_{r}$ recoils in LXe is on the XENON100 
collaboration. Attempts to substitute for this with a biased analysis
represent a lack of consideration for the many efforts made by other 
dark matter researchers working towards similar ends. We invite the XENON100 
collaboration to reconsider their claims, and to include in all
future results a balanced description of the many unknowns and the 
uncertainty they represent.

The authors are indebted to E. Dahl for many useful
comments, and to A. Manzur for the preparation of Fig.\ 1.

\section{Appendix}\label{sec:¥}

Figs.\ 4 and 5 illustrate the dependence on the specific $\mathcal{L}_{\text{eff}}$ 
model used for the extraction of light-WIMP limits from XENON100 data. 
We have followed the procedure described in \cite{xenon100}, allowing the 
sensitivity to be dominated by a 
Poisson tail of photoelectrons generated from recoils from the low-energy region for which 
experimental calibration data do not exist. While we do not condone this 
approach, it allows us to isolate the effect of the choice of 
$\mathcal{L}_{\text{eff}}$ model. Given the very specific meaning of the 
CoGeNT region showed in \cite{xenon100} (which is explicitly mentioned in 
\cite{cogent}), we use experiment regions and astrophysical 
parameters as in Fig.\ 
1 in \cite{andreas} for these calculations. We obtain excellent agreement with the  
sensitivity claimed by XENON100 above $\sim$7 GeV/c$^{2}$, when 
using their flat $\mathcal{L}_{\text{eff}}$ low-energy extrapolation and 
the favorable 
astrophysical parameters employed by them (such as $\rho_{DM}$= 0.5 
GeV/cm$^{3}$ and v$_{esc}$=650 km/s). However, we do not reproduce 
the unexpected power-law behavior of their limit
curve that is evident at lower masses, even when taking into account S1
fluctuations.

To establish a comparison with other models, 
we fit the recent data from Manzur {\it et al.} \cite{dan1} with an adiabatic term as in 
\cite{ahlen}, to account for the expected kinematic cutoff for LXe 
(E$_{max}$). A similar decrease in $\mathcal{L}_{\text{eff}}$ at low 
energies is predicted by the model presented in \cite{dan1}, which 
takes into account the effect of electrons that escape the nuclear 
recoil track, and thus do not recombine to produce scintillation.  ZEPLIN 
data are better fitted by a logistic (sigmoid) function. Both choices 
(adiabatic and logistic) are conservative in the sense that they 
generate a finite amount of scintillation all the way down to zero 
energy, something for which there is no experimental evidence at the present time. 
For the astrophysical parameters used here, 
the adiabatic fit to Manzur {\it et al.}, and a 7 GeV/c$^{2}$ WIMP 
with 5$\times10^{-41}$ cm$^{2}$ spin-independent coupling, we would expect 
0.045 events above a 4 photoelectron threshold for the
XENON100 data set. This becomes 0.25 events for a 3 photoelectron threshold.


\begin{thebibliography}{9}

\bibitem{xenon100}E. Aprile {\it et al.}, {\tt arXiv:1005.0380v1}.
\bibitem{DAMA} R. Bernabei {\it et al.}, Eur.\ Phys.\ 
J.\ {\bf C56} 
(2008) 333.
\bibitem{cogent} C.E. Aalseth {\it et al.}, submitted to Phys. Rev. 
Lett., {\tt arXiv:1002.4703}. 
\bibitem{escape} M.C. Smith {\it et al.}, Mon. Not. Roy. Astron. Soc. 
{\bf 379} (2007) 755, {\tt astro-ph/0611671}. 
\bibitem{dan1}A. Manzur {\it et al.}, Phys. Rev. {\bf C81} 
(2010) 025808, {\tt arXiv:0909.1063}.
\bibitem{zeplin}V.N. Lebedenko {\it et al.}, Phys. Rev. {\bf D80} (2009) 052010,{\tt arXiv:0812.1150}.
\bibitem{theiraps} A. Manzur (XENON10 collaboration), presented at the 
2007 APS April meeting, available from\\ http://xenon.astro.columbia.edu/presentations.html
\bibitem{peter}P. Sorensen (XENON10 collaboration), presented at the 
2010 Light Dark matter Workshop, UC Davis, available from
http://particle.physics.ucdavis.edu/seminars/  \\ 
data/media/2010/apr/sorensen.pdf
\bibitem{prevnim}P. Sorensen {\it et al.}, Nucl. Instr. Meth. {\bf 
A601} 
(2009) 339.
\bibitem{ahlen}D.J. Ficenec {\it et al.}, Phys. Rev. {\bf D36} 
(1987) R311; S.P. Ahlen {\it et al.}, Phys. Rev. {\bf D27} 
(1983) 688.
\bibitem{review}V.I. Tretyak, Astropart. Phys. {\bf 33} 
(2010) 40.
\bibitem{spooner}J. Hong {\it et al.}, Astropart. Phys. {\bf 16} 
(2002) 333.
\bibitem{doke}T. Doke {\it et al.}, Jpn. J. Appl. Phys. {\bf 41} 
(2002) 1538.
\bibitem{aprile}E. Aprile {\it et al.}, Phys. Rev. {\bf D72} 
(2005) 072006.
\bibitem{ziegler}J.F. Ziegler, J. Appl. Phys / Rev. Appl. Phys. {\bf 
85} (1999) 1249.
\bibitem{lumin}R.H. Bartram {\it et al.}, J. Luminesc. {\bf 68} 
(1996) 225.
\bibitem{wonder}R.F. Lang, priv. comm.; E. Aprile, presented at 
WONDER 2010, Gran Sasso, available from http://wonder.lngs.infn.it/
\bibitem{jcapetc}P.S. Barbeau {\it et al.}, Nucl.\ 
Instr.\ 
Meth.\ {\bf A574} (2007) 385; P.S. Barbeau {\it et al.}, JCAP {\bf 09}
(2007) 
009; P.S. Barbeau, Ph.D. Diss., University of Chicago (2009).
\bibitem{tex} A review of germanium quenching factor measurements can 
be found in S.T. Lin {\it et al.}, Phys. Rev. {\bf D79} (2009) 
061101, {\tt arXiv:0712.1645}.
\bibitem{andreas}S. Andreas {\it et al.}, {\tt arXiv:1003.2595}.
\bibitem{gondolo}P. Gondolo, presented at the 
2010 Light Dark matter Workshop, UC Davis, available from
http://particle.physics.ucdavis.edu/seminars/  \\ 
data/media/2010/apr/gondolo.pdf
\end{thebibliography}
\end{document}